\documentclass[prb,aps,twocolumn,showpacs,superscriptaddress,twoside]{revtex4-1}
\usepackage{amsmath}
\usepackage{amssymb}
\usepackage{graphicx,color}

\begin{document}

\title{Surface scattering and band gaps in rough waveguides and nanowires}

\author{O.~Dietz}
\affiliation{Fachbereich Physik, Philipps-Universit\"{a}t Marburg, Germany}
\affiliation{Institut f\"{u}r Physik, Humboldt-Universit\"{a}t zu Berlin, Germany}
\email{otto.dietz@physik.hu-berlin.de}

\author{H.-J.~St\"{o}ckmann}
\affiliation{Fachbereich Physik, Philipps-Universit\"{a}t Marburg, Germany}

\author{U.~Kuhl}
\affiliation{Fachbereich Physik, Philipps-Universit\"{a}t Marburg, Germany}
\affiliation{LPMC, CNRS UMR 7336, Universit\'{e} de Nice Sophia-Antipolis, 06108 Nice, France}
\email{ulrich.kuhl@unice.fr}

\author{F.~M.~Izrailev}
\affiliation{Instituto de F\'{\i}sica, Universidad Aut\'{o}noma de Puebla, Mexico}
\affiliation{NSCL and Department of Physics and Astronomy, Michigan State
University, E.~Lansing, Michigan 48824-1321, USA}

\author{N.~M.~Makarov}
\affiliation{Instituto de Ciencias, Universidad Aut\'{o}noma de Puebla, Puebla, Mexico}

\author{J.~Doppler}
\affiliation{Institute for Theoretical Physics, Vienna University of Technology, A-1040 Vienna, Austria, EU}
\author{F.~Libisch}
\affiliation{Institute for Theoretical Physics, Vienna University of Technology, A-1040 Vienna, Austria, EU}
\author{S.~Rotter}
\affiliation{Institute for Theoretical Physics, Vienna University of Technology, A-1040 Vienna, Austria, EU}

\date{\today}

\begin{abstract}
The boundaries of waveguides and nanowires have drastic influence on their coherent scattering properties. Designing the boundary profile is thus a promising approach for transmission and band-gap engineering with many applications. By performing an experimental study of microwave transmission through rough waveguides we demonstrate that a recently proposed surface scattering theory can be employed to predict the measured transmission properties from the boundary profiles and vice versa. A new key ingredient of this theory is a scattering mechanism which depends on the squared gradient of the surface profiles. We demonstrate the non-trivial effects of this scattering mechanism by detailed mode-resolved microwave measurements and numerical simulations.
\end{abstract}


\maketitle


The coherent scattering through systems with surface roughness is a ubiquitous phenomenon which occurs on vastly different length and time scales.\cite{mar07,bas79} The effects induced by surface scattering often are the key for the understanding both of natural phenomena, like the scattering of underwater waves at a rough ocean seabed,\cite{fri94} as well as of man-made devices like optical fibers and waveguides, \cite{cha05,pha09,lee00,mak11} photonic crystal devices,\cite{chu99,rob05,gan12} metamaterials,\cite{mar11} thin metallic films,\cite{fis89c,mey99} layered structures,\cite{zha92,gam99} nanowires,\cite{hub08,akg08,fei06,*fei09} in optical diffraction tomographs,\cite{mai09} and confined quantum systems in general.\cite{fer09} For some devices, like quantum cascade lasers\cite{khu09} and gated graphene nano-ribbons\cite{han07,muc09,eva08} the scattering at rough boundaries was identified as one of the dominant factors that limits the device performance. Surface roughness effects might hold the key for the explanation of anomalously large persistent currents in metallic rings\cite{fei12} and are actively used to control gravitationally bound quantum states of neutrons\cite{nes02,jen11} as well as to enhance the thermo-electric performance of nano-wires.\cite{hoc08} The understanding of all these phenomena rests on a predictive surface scattering theory that relates the properties of a rough surface to the transmission characteristics of the corresponding device and vice versa.

It is here that our analysis sets in to show that conventional surface scattering theories for the coherent transmission through waveguides lack significant ingredients. If the aim is to determine the transmission properties of a system from a given rough boundary profile, numerical calculations can sometimes help to overcome these deficiencies of existing analytical methods.\cite{akg08,zho10} If, however, the aim is to solve the inverse problem, i.e., to design a boundary such as to obtain desired transmission properties, improved analytical models are indispensable which provide the functional dependence of the transmission properties on the boundary profile. It is this important missing link that has recently been provided in terms of an analytical surface scattering theory (SST),\cite{izr05b,izr05c,ren11a,ren11b} which we test here for the first time.

This new SST predicts how mode-specific scattering lengths in waveguides depend on the details of a system's surface roughness. It introduces the previously neglected \textit{square-gradient scattering} (SGS) mechanism and predicts that this new scattering mechanism has to be considered together with the conventional \textit{amplitude scattering} (AS) mechanism.\cite{bas79} Although SGS is related to higher-order terms in the disorder strength $\sigma$ it can even be the major scattering mechanism in systems with modest disorder. SST focuses, however, on long-range correlations, which seems to restrict its applicability to very \textit{long waveguides}. In this Rapid Communication we resolve this impasse by demonstrating that the validity of SST extends to \textit{short, individual} waveguides as well. We thereby provide the key for an accurate description of transmission through surface-disordered waveguides in general and a solid basis for a deeper understanding of many of the applications mentioned above. In addition, our results allow us to impose desired transmission properties on a waveguide through the roughness of its boundaries.

\begin{figure}
\includegraphics[width=\columnwidth]{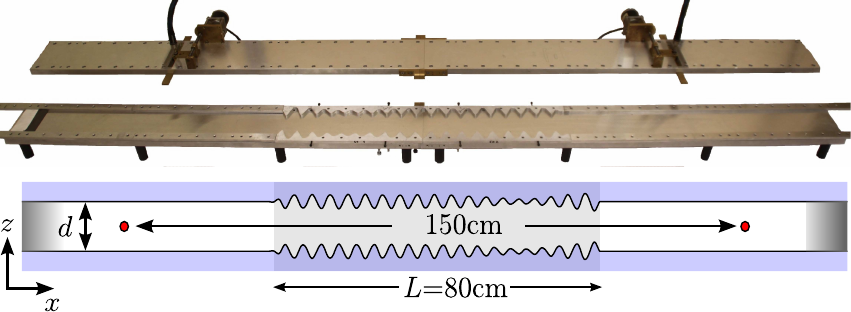}
\caption{\label{fig:boundary} (Color online)
Experimental setup with exchangeable rough boundaries. The photo shows the considered waveguide with the surface disordered region in the center and the lifted top plate. Step motors are attached to the top plate to shift the input and output antenna in $z$ direction. Absorbers at both ends suppress reflection (indicated  in the sketch below).}
\end{figure}

To demonstrate this explicitly we perform a proof-of-principle experiment using the microwave setup shown in Fig.~\ref{fig:boundary}. The studied scattering system consists of a planar waveguide with metallic walls with a ``rough'' central part of length $L=80$\,cm which features two undulating, symmetric boundaries, $z=\pm[d/2+\sigma\xi(x)]$. The average width of the scattering part is $d=10$\,cm and the roughness variance is $\sigma^2$. In the experiment we choose a modest disorder of $\sigma/d \approx 9\%$ ($\sigma\approx8.7$\,mm). The boundary profile function $\xi(x)$ has zero mean, $\langle\xi(x)\rangle=0$, unit variance, $\langle\xi^2(x)\rangle=1$, and features a binary correlator defined by
\begin{equation}\label{eq:<xixi>}
\langle\xi(x)\,\xi(x')\rangle={\cal W}(x-x').
\end{equation}
In the convention of the SST the angular brackets stand for an ensemble average over different realizations of $\xi(x)$. However, by assuming the ergodicity of $\xi(x)$, the average can be performed along a single surface profile. Therefore, the quantity ${\cal W}(x)$ can also be treated as the autocorrelation function of the boundary profile used in the experiment.

For vanishing disorder the wave vector is quantized along the transverse direction ($z$), yielding $n$ propagating modes whose wave numbers in guiding direction ($x$) at frequency $\nu$ are given by
\begin{equation}\label{eq:kn}
k_{n}=\sqrt{\left(\frac{2\pi\nu}{c}\right)^2-\left(\frac{\pi n}{d}\right)^2}\,.
\end{equation}
The chosen up-down symmetry of the disorder profile prevents mixing of odd and even modes during the scattering process. We also restrict our study to the frequency range in which we can assume single-mode scattering for the odd and even modes.

The localization length $L^\textrm{(loc)}_n$ is defined via the famous expression for the average logarithm of the transmittance, $\langle\ln T_{n}\rangle=-2L/L^\textrm{(loc)}_n$, showing an exponential decrease with the sample length $L$. In the present experiment we deal with a single, finite boundary realization. Thus the partial transmittances $T_{n}$ correspond to the average over the most probable (representative) realizations of the surface disorder,\cite{izr12}
\begin{equation} \label{eq:Tn}
\langle T_{n}\rangle=\exp\left(-2L/L^\textrm{(loc)}_n\right).
\end{equation}
For isolated modes the localization length $L^\textrm{(loc)}_n$ of the $n$th propagating mode ($n=1,2$) is twice the corresponding backscattering length, $L^\textrm{(loc)}_n=2L^{(b)}_n$, a quantity which has been calculated with SST,\cite{ren11a}
\begin{equation}\label{eq:Ln}
\frac{1}{L^{(b)}_n}=\frac{1}{L_n^{(b),(AS)}}+\frac{1}{L_n^{(b),(SGS)}}\,.
\end{equation}
Here, the two different terms originate from the amplitude scattering (AS) and the square-gradient scattering (SGS) mechanisms, respectively:
\begin{eqnarray}
\frac{1}{L_n^{(b),(AS)}}&=&\frac{\sigma^2}{d^6}\frac{4\pi^4n^4}{k_n^2}\,W(2k_n)\,,\label{eq:Ln-AS}\\
\frac{1}{L_n^{(b),(SGS)}}&=&\frac{\sigma^4}{d^4}\frac{(3+\pi^2n^2)^2}{18k_n^2}\,S(2k_n)\label{eq:Ln-SGS}\,.
\end{eqnarray}
Correspondingly, these transmission characteristics are specified by two different roughness power spectra,\cite{ren11a}
\begin{eqnarray}
W(k_x)&=&\int_{-\infty}^\infty{\cal W}(x)\exp(-ik_xx)\mathrm{d}x\,,
\label{eq:W}\\
S(k_x)&=&\int_{-\infty}^\infty{\cal W}''^2(x)\exp(-ik_xx)\mathrm{d}x\,.
\label{eq:S}
\end{eqnarray}
The first one, $W(k_x)$, is the Fourier transform of the roughness-height correlator \eqref{eq:<xixi>}, and is therefore called \textit{roughness-height} power spectrum. The second term, $S(k_x)$, is referred to as the \textit{roughness-square-gradient} power spectrum since for gaussian random processes it is the Fourier transform of the roughness-square-gradient correlator, $\langle\xi'^2(x)\,\xi'^2(x')\rangle-\langle\xi'^2(x)\rangle^2=2{\cal W}''^2(x-x')$.\cite{izr05b,ren07a}

In order to observe the two scattering mechanisms separately, we generate a rough boundary profile $\xi(x)$ with a power spectrum $W(2k_n)$ being non-zero only within a small $k_n$-range. Then the localization length and potential transmission gaps outside this range will be determined entirely by $S(2k_n)$. For this purpose we choose the roughness-height power spectrum in the following rectangular form:
\begin{equation}\label{eq:WOD-W}
W(k_x)=\left\{\begin{array}{cl}\frac{\pi}{2(k_{+}-k_{-})},&\textrm{for}\quad 2k_{-}<|k_x|<2k_{+}\\0,&\textrm{otherwise}.\end{array}\right.,
\end{equation}
respecting the normalization ${\cal W}(0)=1$. In our experiment we specify the correlation parameters $k_{\mp}$ as follows: $k_{-}=70$\,m$^{-1}$ and $k_{+}=80$\,m$^{-1}$. Transmission properties of rectangular spectra \eqref{eq:WOD-W} have been investigated theoretically\cite{ren11b,izr12} and tested in microwave experiments\cite{kuh08a,die11a} in one-dimensional and quasi-one-dimensional disordered systems with discrete scatterers.

The disorder $\xi(x)$ with a desired power spectrum $W(k_x)$ was constructed with a widely used convolution method (see, e.\,g., Ref.~\onlinecite{izr12} and references therein). First, starting from a prescribed power spectrum $W(k_x)$, we derive the modulation function $G(x)$ whose Fourier transform is $\sqrt{W(k_x)}$,
\begin{equation}\label{eq:MF}
G(x)=\int_{-\infty}^{\infty}\frac{dk_x}{2\pi}\,\sqrt{W(k_x)}\exp\left(ik_xx\right).
\end{equation}
In the second step the disordered profile $\xi(x)$ is generated as a convolution of a $\delta$-correlated random process $\alpha(x)$, with $\langle\alpha(x)\rangle=0$, and the modulation function $G(x)$,
\begin{equation}\label{eq:xiG}
\xi(x)=\int_{-\infty}^\infty\,dx'\,\alpha(x-x')\,G(x').
\end{equation}
From Eqs.~\eqref{eq:MF} and \eqref{eq:xiG}, one can readily recognize that the algorithm for constructing a rough surface $\xi(x)$ with power spectrum \eqref{eq:WOD-W} reads,\cite{ren11b,izr12}
\begin{equation}\label{eq:xi-WOD}
\xi(x)=\int_{-\infty}^\infty\,\frac{dx'}{\sqrt{2\pi}}\,\alpha(x-x')\,\frac{\sin(2k_+x')-\sin(2k_-x')}{(k_+-k_-)^{1/2}x'}.
\end{equation}
Instead of a continuous $\alpha(x)$ we used a discrete series of random numbers to generate 60 discrete $\xi(x)$-values at $x=mL/60$ ($m=1,2,3,\ldots,60$). The continuous roughness $\xi(x)$ was recovered by spline interpolation. After such a procedure the resulting boundary profile is no longer a pure gaussian random process and its power spectrum $W(k_x)$ is not perfectly rectangular as in Eq.~\eqref{eq:WOD-W}. However, it still approximates the peaked structure with the same $k_x$-range. Note that the boundary has mainly two scales: $2\pi/(k_{+}+k_{-})$ (fast oscillations) and $2\pi/(k_{+}-k_{-})$ (beating oscillation) which can both be seen in Fig.~\ref{fig:boundary}.

\begin{figure}
\includegraphics[width=.8\columnwidth]{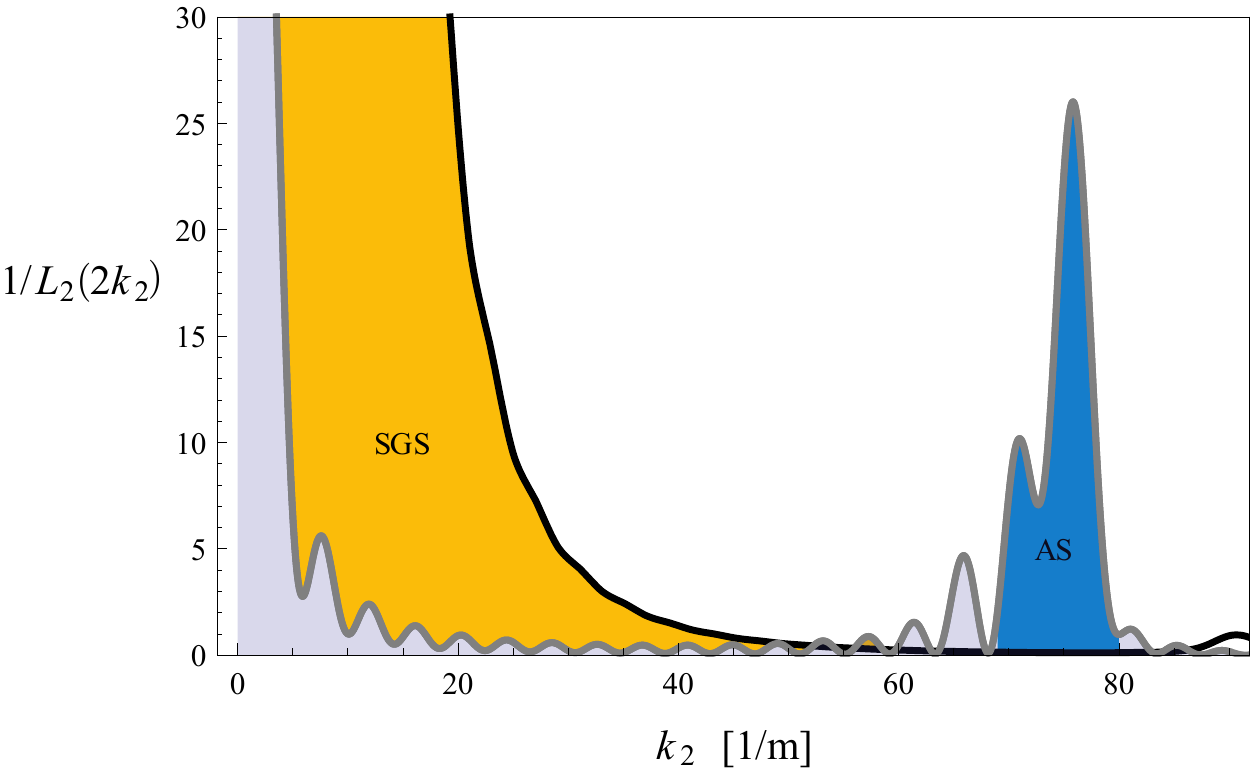}
\caption{\label{fig:vANDw} (Color online)
The inverse backscattering lengths $1/L_2^{(b),(AS)}$ (gray curve) and $1/L_2^{(b),(SGS)}$ (black curve) for the second mode, $n=2$ are shown; see Eqs.~\eqref{eq:Ln-AS} and \eqref{eq:Ln-SGS}. The corresponding power spectra are calculated directly from the rough boundaries shown in Fig.~\ref{fig:boundary}. Regions of dominating SGS and AS mechanisms are shaded.}
\end{figure}

The inverse backscattering lengths, $1/L_n^{(b),(AS)}$ and $1/L_n^{(b),(SGS)}$, following from SST for the constructed boundary profile are shown in Fig.~\ref{fig:vANDw} for the second mode $n$=2 (similar behavior for the first mode $n=1$ is not shown). We see that the inverse AS backscattering length \eqref{eq:Ln-AS} shows the expected peak at $k_n\approx75$\,m$^{-1}$, determined by $W(2k_n)$, whereas the corresponding curve for the inverse SGS backscattering length \eqref{eq:Ln-SGS} dominates at much lower frequencies due to the influence of $S(2k_n)$. Note that as $k_n\to0$ both curves in Fig.~\ref{fig:vANDw} diverge due to the factor $k_n^{-2}$. However, these divergencies are characteristically different for small frequencies. For the AS mechanism the factor $W(2k_n)$ vanishes, $\lim_{k_n\to0}W(2k_n)=0$, while for the SGS mechanism the factor $S(2k_n)$ remains finite, $\lim_{k_n\to0}S(2k_n)=8\pi(k_{+}^5-k_{-}^5)/5(k_{+}-k_{-})^2>0$. We conclude that the gap in the waveguide transmission for small values of $k_n$ is thus primarily induced by the SGS mechanism, unlike the transmission gap at around $k_n=75$\,m$^{-1}$ which is associated with the AS mechanism.

In Ref.~\onlinecite{die11a} we have demonstrated that the type of experimental setup we use (see Fig.~\ref{fig:boundary}) allows for measurements of the scattering matrix $S_{nm}$ (in the mode-representation). Here we focus on the squared modulus of the diagonal elements of the scattering matrix which are proportional to the transmittance \eqref{eq:Tn},\cite{ren11a,die11a}
\begin{equation}\label{eq:Snn}
\langle T_{n}\rangle=c_n|S_{nn}|^2.
\end{equation}
The properties of the antennas and their coupling to the waveguide modes are captured in the coefficients $c_n$. In the experiment they are not directly accessible, but can be determined by fitting the transmittance through the corresponding mode for a roughness-free waveguide. A theoretical study of antenna coupling\cite{tud08} predicts that the $c_n$ depend only weakly on frequency such that we can regard them as fitting constants. Due to an imperfect coupling to the antennas and losses at open ends, the maximal measurable transmission is of the order of $10^{-3}$ as reflected in the values of $c_1=6.33 \times 10^{3}$ and $c_2=3.33 \times 10^{3}$ (for details see Ref.~\onlinecite{die11a}).

\begin{figure}
\includegraphics[width=\columnwidth]{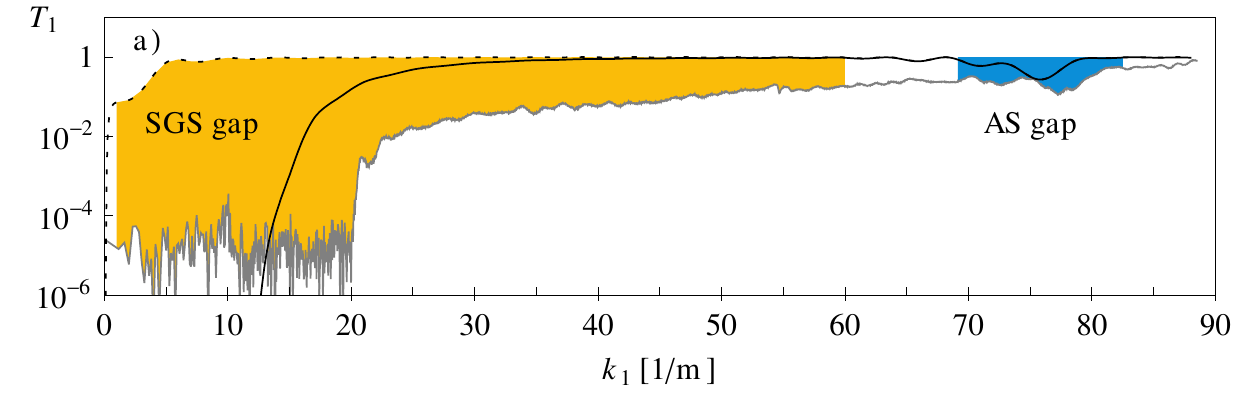}
\includegraphics[width=\columnwidth]{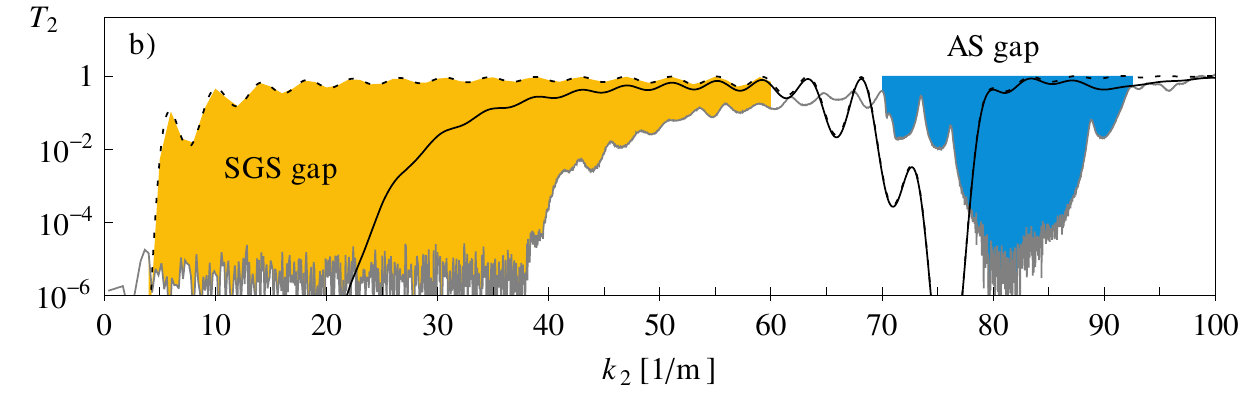}
\includegraphics[width=\columnwidth]{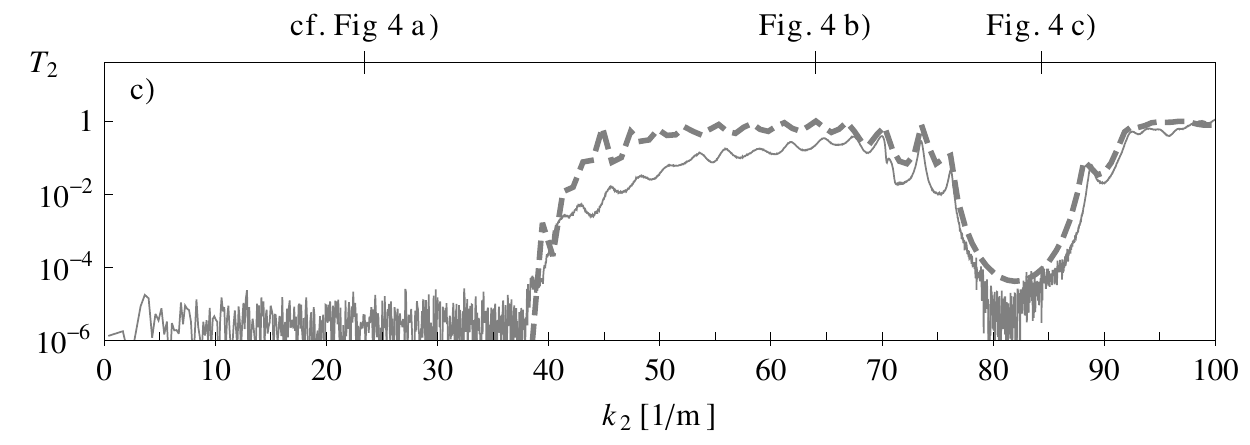}
\includegraphics[width=\columnwidth]{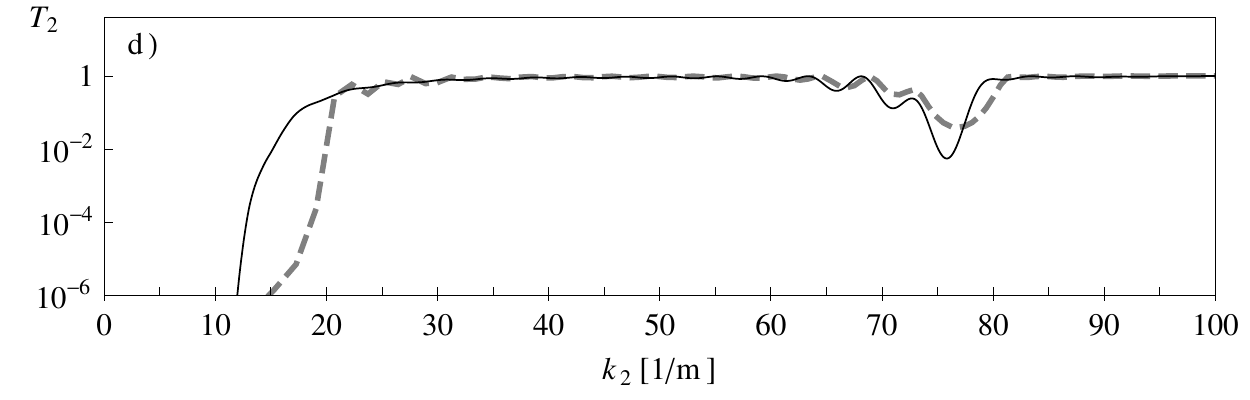}
\caption{\label{fig:results} (Color online)
Measured transmittance $T_{n}$ (gray curve), for (a) the first ($n=1$) and (b) the second mode ($n=2$). The corresponding theoretical prediction is shown as solid black curve [see Eq.~\eqref{eq:Tn}]. For the dotted line the SGS is neglected explicitly. The shaded areas indicate the gaps caused by the dominating scattering mechanism. (c) Numerically calculated transmittance for the second mode $T_{2}$ (dashed gray line), compared to measured transmittance (solid gray). The markers indicate the positions of the numerically calculated wave functions shown in Fig.~\ref{fig:numerical}. (d) Numerically calculated transmittance (dashed gray curve) compared to the theoretical prediction (black solid curve) for reduced disorder strength $\sigma$ (50\% of experimental value).}
\end{figure}

The resulting transmission curves as measured for the first and second mode feature two transmission gaps [see gray curves in Figs.~\ref{fig:results}(a) and \ref{fig:results}(b), respectively]. As predicted by SST (see solid black curves for comparison), the first transmission gap is located near the opening of the modes and the second gap appears around $k_n=80$\,m$^{-1}$. This clear correspondence between the experiment and theory allows us to associate the observed gaps with the SGS and AS mechanisms, respectively. To further prove this association we also calculate theoretical curves where the SGS term is explicitly neglected [see dashed curves in Figs.~\ref{fig:results}(a) and \ref{fig:results}(b)]. These curves only contain the AS terms and, indeed, do not display the previously identified SGS-gaps. We thus arrive at the central message of our Rapid Communication: transmission through surface-disordered waveguides manifests clear signatures of the AS and the SGS mechanism. The SGS contribution to transmission may even dominate the conventional AS contribution and both mechanisms can introduce extended band gaps in the transmission.

\begin{figure}
\includegraphics[width=.95\columnwidth]{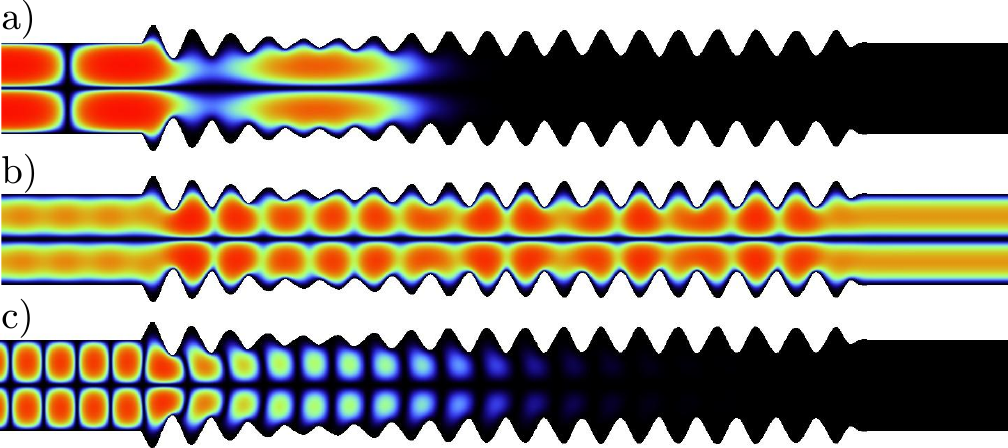}
\caption{\label{fig:numerical}(Color online)
Numerically calculated wave functions for transmission in the second mode. The corresponding $k_2$-values of (a) $23.46$\,m$^{-1}$, (b) $64.02$\,m$^{-1}$, and (c) $84.39$\,m$^{-1}$ are indicated in Fig.\,\ref{fig:results}(c).}
\end{figure}

SST thus reproduces all experimental features \textit{qualitatively} correctly. There are, however, systematic \textit{quantitative} discrepancies between experiment and theory. Most prominently, the frequency interval of the observed transmission gaps are shifted to higher frequencies than those predicted [see Fig.~\ref{fig:results}(a) and \ref{fig:results}(b)]. In particular for the second mode this shift is quite sizable. To check the origin of these discrepancies we performed detailed numerical simulations for the same scattering setup. We calculated all scattering matrix elements and the corresponding scattering wave functions (see Fig.~\ref{fig:numerical}) in the waveguide with an extended version of the modular Green's-function method\cite{rot00b,libisch11a} which is based on a finite-difference approximation of the two-dimensional scattering geometry. Both the AS gap and the SGS gap for the numerical transmittance $T_{2}$ match the experimental values [see dashed gray curve in Fig.~\ref{fig:results}(c)]. We thus conclude that the observed shifts and broadenings of the gaps are not experimental artifacts, but rather details omitted by SST. Since the latter was derived for perturbatively small disorder amplitude $\sigma$, an obvious source of the theoretical shortcomings may be the sizable value of $\sigma$ in our experiment which goes beyond the first-order correction inherent in the AS mechanism as well as the second-order correction in the SGS mechanism. To test this assumption we compare the numerical calculations and the corresponding theoretical predictions for a waveguide system with the same boundary profile, but with half the roughness value of $\sigma$ [see Fig.~\ref{fig:results}(d)]. In this case we, indeed, obtain much better agreement between theory and numerics: The AS gap is now both less broad and shifted to the predicted interval (note that gap positions do not depend on $\sigma$ in SST) and the SGS gap is suppressed, as its dependence $\sim\sigma^4$ in Eq.~\eqref{eq:Ln-SGS} would suggest.

Our numerics also allow us to visualize the scattering wave functions at characteristic frequency values (see Fig.~\ref{fig:numerical}): at $k_2=23.46$m$^{-1}$ the second mode is fully backscattered due to the SGS (a). Once this SGS gap is passed full transmission is recovered (b), followed by the AS-gap (c) where complete backscattering is found.

To summarize, we investigated coherent transmission through a quasi-one-dimensional waveguide with surface disorder. Using a microwave measurement setup we find pronounced transmission gaps in predetermined frequency intervals, each of which we can associate with a specific surface scattering mechanism, i.e., the \textit{amplitude scattering} and the \textit{square-gradient scattering}. An observed shift of the amplitude scattering gap could be attributed to the non-vanishing disorder strength. Our investigations show that even relatively short waveguides can exhibit effects predicted for systems with long-range correlations leading to drastic changes in their transmission properties. This result opens up the way to design boundary profiles to induce desired transmission properties in surface-disordered scattering systems.

Support by the DFG within the research group 760 ``Scattering Systems with Complex Dynamics'' is acknowledged by O.D., H.-J.S., and U.K. F.M.I. and N.M.M. gratefully acknowledge the support of the SEP-CONACyT (Mexico) Grants No.~80715 and No.~166382. J.D., F.L., and S.R. acknowledge support by the Vienna Science and Technology Fund (WWTF) through Project No. MA09-030, by the Austrian Science Fund (FWF) through Project No. P14 in the SFB IR-ON, and computational resources by the Vienna Scientific Cluster (VSC).

\end{document}